\documentclass[aasms,tighten,psfig,graphicx,11pt,a4]{aastex}
\slugcomment{Submitted to {\it Ap. J. Letters}, accepted 31 Jan 2000}

\begin{document}

\title{Water Ice in 2060 Chiron and its Implications for Centaurs
and Kuiper Belt Objects}
\author{Jane X. Luu}
\affil{Sterrewacht Leiden\\
Postbus 9513,
2300RA Leiden, The Netherlands}

\author{David C. Jewitt\altaffilmark{1}}
\affil{Institute for Astronomy\\
2680 Woodlawn Drive, Honolulu, HI 96822}

\and

\author{Chad Trujillo\altaffilmark{1}}
\affil{Institute for Astronomy\\
2680 Woodlawn Drive, Honolulu, HI 96822}

\altaffiltext{1}{Visiting Astronomer, W. M. Keck Observatory,
jointly operated by California Institute of Technology and the University
of  California.}

\newpage

\begin{abstract}

We report the detection of water ice in the Centaur 2060 Chiron, based
on near-infrared spectra (1.0 - 2.5 $\micron$) taken with the
3.8-meter United Kingdom Infrared Telescope (UKIRT) and the 10-meter
Keck Telescope.  The appearance of this ice is correlated with the
recent decline in Chiron's cometary activity: the decrease in the coma
cross-section allows previously hidden solid-state surface features to
be seen.  We predict that water ice is ubiquitous among Centaurs and
Kuiper Belt objects, but its surface coverage varies from object to
object, and thus determines its detectability and the occurrence of cometary
activity.

\end{abstract}


\newpage

\section{Introduction}

The Centaurs are a set of solar system objects whose orbits are
confined between those of Jupiter and Neptune.  Their planet-crossing
orbits imply a short dynamical lifetime ($10^6
- 10^7$ yr).  The current belief is that Centaurs are objects
scattered from the Kuiper Belt that may eventually end up in the inner
solar system as short-period comets.  The first discovered and
brightest known Centaur, 2060 Chiron, is relatively well studied.  The
object is firmly established as a comet, with a weak but persistent
coma.  It is well documented that Chiron possesses neutral colors
(e.g., Hartmann et al. 1990, Luu and Jewitt 1990), and a low albedo of
$0.14^{+0.06}_{-0.03}$ (Campins et al. 1994).  (It must be noted that
most of these measurements were made when Chiron clearly exhibited a
coma so that the measurements are likely to have been contaminated by
dust scattering from the coma.  The albedo, in particular, should be
viewed as an upper limit).  Chiron has a rotation period of $\sim 6$
hr (e.g., Bus et al. 1989) and a photometric amplitude that is
modulated by the cometary activity level (Luu and Jewitt 1990,
Marcialis and Buratti 1993, Lazzaro et al. 1997).  Published optical
and near-IR spectra of Chiron show a nearly solar spectrum, varying from
slightly blue to completely neutral (Hartmann et al. 1990, Luu 1993,
Luu et al. 1994, Davies et al. 1998), and devoid of specific
mineralogical features.

As a group, the Centaurs display remarkable spectral diversity (e.g.,
Luu and Jewitt 1996).  The Centaur 5145 Pholus is among the reddest bodies
in the solar system (Mueller et al. 1992, Fink et al. 1992), and shows
absorption features at 2.00 and 2.25 $\micron$ (see, e.g., Davies et
al. 1993a, Luu et al. 1994).  Cruikshank et al. (1998) interpreted the
2.0 $\micron$ feature as due to water ice, and the 2.27 $\micron$ due
to methanol.  They derived a best fit to the spectrum which consisted
of carbon black and an olivine-tholin-water-methanol mixture.
Pholus's extreme red color and low albedo ($0.044 \pm 0.013$, Davies
et al. 1993b) strongly suggest that long-term irradiation of carbon-
and nitrogen-bearing ices has resulted in an organic-rich, dark
"irradiation mantle" (e.g., Johnson et al. 1987).  The spectral
differences between Chiron and Pholus have been attributed to
the presence of cometary activity in Chiron and the lack thereof in
Pholus (Luu 1993, Luu et al. 1994).  A continuous rain of sub-orbital
cometary debris falling onto the surface of Chiron may have buried a
more primordial irradiation mantle with unirradiated matter ejected
from the interior.  In this paper we show further evidence supporting
this hypothesis, and that scattering effects by coma dust particles
also eliminate spectral features seen in solid surface reflection.

\section{Observations}

\noindent {\bf (a) Keck observations}. The Keck near-infrared
observations were made on UT 1999 April 03, at the $f$/25 Cassegrain
focus of the Keck I telescope, using the NIRC camera (Matthews and
Soifer 1994).  The NIRC detector is a 256 x 256 pixel InSb array that
can be switched from direct imaging to slit spectroscopy.  In imaging
mode the pixel scale is 0.15 arcsec per pixel (38" x 38" field of
view), and in spectroscopy mode the resolution is $\lambda /
\Delta\lambda \approx 100$.  We used a 0.68" $\times$ 38" north-south
slit for all spectral observations.  Uneven illumination of the slit
and pixel-to-pixel variations were corrected with spectral flat fields
obtained from a diffusely illuminated spot inside the dome.

Since the target was not visible during spectral observations,
Chiron's position was confirmed by centering it at the location of the
slit and taking an image.  The slit, grism, and blocking filter were
then inserted in the beam for the spectroscopic observation.
Spectra were made in pairs dithered along the slit by 13".  We
obtained spectra in two different grating positions, covering the JH
wavelength region (1.00 - 1.50 $\micron$) and the HK region (1.4 -
2.5 $\micron$).  Non-sidereal tracking at Keck showed a slight drift
with time, so we recentered Chiron in the slit every 15-20 minutes.

\noindent {\bf (b) UKIRT observations}. The UKIRT observations were
made on UT 1996 Feb 7 and 8 with the CGS4 infrared spectrometer
mounted at the Cassegrain focus.  The detector was a 256 $\times$ 256
pixel InSb array, with a 1.2" per pixel scale in the spatial
direction.  An optical TV camera fed by a dichroic beam splitter gave
us slit viewing capability and thus we were able to guide on the
target during all observations.  The conditions were photometric and
the image quality was $\sim$ 1" Full Width at Half Max (FWHM), so we
used a 1.2'' $\times$ 80'' slit aligned North-South on the sky at all
times.  A 75 line per mm grating was used in first order for all
observations, yielding a dispersion of 0.0026 $\micron$/pixel in the H
and K band ($\lambda / \Delta\lambda \approx 850$).  However, the
detector was dithered by 1/2 pixel during each observation, so the
spectra were oversampled by a factor of 2.  The effective spectral
coverage in the H band was 1.4 -- 2.0 $\micron$ and in the K band 1.9
-- 2.4 $\micron$.  Sky background removal was achieved by nodding the
telescope 30" (23 pixels) along the slit.  Dark frames and calibration
spectra of flat fields and comparison lamps (Ar) were also taken every
night.

Both the Keck and UKIRT observations were calibrated using stars on the
UKIRT Faint Standards list (Casali and Hawarden 1992).  At both
telescopes, we took care to observe the standard stars at airmasses
similar to those of Chiron (airmass difference $\leq$
0.10), to ensure proper cancellation of sky lines. 

The separate reflectance spectra from each night of observation are
shown in Fig. 1.

\section{Discussion}

\subsection{The spectra}

The Chiron spectra (Fig. 1) show that: (a) Chiron is nearly neutral in
the 1.0 -- 2.5 $\micron$ region; (b) there is a subtle but definite
absorption feature at 2 $\micron$ ($\sim$ 0.35 $\micron$ wide, $\sim
10\%$ deep) in spectra from 1996 and 1999, and a marginal absorption
feature near 1.5 $\micron$ in the 1999 spectrum; and (c) the spectral
slope and the strength of the 2 $\micron$ feature change with time.

Reflectivity gradients in the JHK region span the range $S'$ = -2
\%/1000~\AA\ to $S' =$ 1\%/1000~\AA\ (Table 2).  In Fig. 2 we compare
Chiron spectra from 1993 (from Luu et al. 1994) with the present
observations.  The flat and featureless 2 $\micron$ spectrum from 1993
stands in sharp contrast with the later spectra.  The presence of the
2 $\micron$ feature in different spectra taken with different
telescopes, instruments, and spectral resolutions provides convincing
evidence that it is real.

The 2 $\micron$ and 1.5 $\micron$ features are clear signatures of
water ice, and the shallowness of the features (compared to that of
pure water ice) indicates that this ice is mixed with dark impurities
(see Clark and Lucey 1984).  We note that the Chiron spectra are very
similar to spectra of minerals and water ice (compare the spectra in
Fig. 2 with Fig. 14 and 15 of Clark 1981, respectively). In Fig. 3 we
show that the Keck Chiron spectrum is well fitted by a model
consisting of a linear superposition of a water ice spectrum and an
olivine spectrum.  The olivine spectrum is needed to provide the
required continuum slope, and at the very small grain size used, the
spectrum of olivine is essentially featureless.  The water ice
spectrum was calculated based on the Hapke theory for diffuse
reflectance (Hapke 1993) and used a grain diameter of 1 $\micron$.  A
description can be found in Roush (1994).  However, we caution that
the model is non-unique, and due to the many free parameters in the
model (e.g., grain albedo, porosity, roughness), the $1 \micron$ grain
size should not be taken literally.  Similarly, olivine could also be
replaced in the fit by other moderately red, featureless absorbers.

The 2 $\micron$ feature in Chiron is clearly time-variable: it was not
apparent in 1993 but changed to a depth of 8 - 10\% in 1996
and 1999.  Chiron's lightcurve variations can be
explained by the dilution of the lightcurve by an optically thin coma
(Luu and Jewitt 1990).  Based on this model, we estimate that the 1993
coma cross-section was $\sim 1.5$ times larger than in 1996 or now.
The likeliest explanation for the time-variability of the $2 \micron$
feature is the degree of cometary activity in Chiron.  In 1993,
Chiron's activity level was high (Luu and Jewitt 1993, Lazzaro et
al. 1997), resulting in a featureless spectrum dominated by scattering
from the coma. By 1996, when the UKIRT observations were made,
Chiron's total brightness had dropped by $\sim 1$ mag to a minimum level
(comparable to that of 1983--1985, Lazzaro et al. 1997), leaving 
spectral contamination by dust at a minimum.

\subsection{Implications of water ice on Chiron}

\subsubsection{Cometary activity in Centaurs}

Considering (1) Chiron's time-variable spectrum, (2) the
presence of surface water ice, and (3) Chiron's persistent cometary
activity, we conclude that Chiron's surface coverage is {\it not}
dominated by an irradiated mantle but more probably by a layer of
cometary debris.  Occultation observations suggest that
sublimation by supervolatiles (e.g., CO, N$_2$) on Chiron occurs in a
few localized icy areas (Bus et al. 1996). Dust grains ejected at
speeds $< 100$ m s$^{-1}$ (the escape velocity) will reimpact the
surface, building a refractory layer which tends to quench sublimation.
Nevertheless, the outgassing is still sustained by the sporadic
exposure of fresh ice on the surface.  Sublimation experiments with
cometary analogs illustrate this phenomenon: outgassing produces a
dust layer, but fresh icy material can still be periodically exposed
by avalanches and new vents created by impacts from large dust
particles (Gr\"un et al.  1993). If, in keeping with a Kuiper Belt
origin, Chiron once possessed an irradiation mantle, we suspect that
it has been either blown off by sublimation or buried under a dust
layer thick enough to mask its features.  If so, the present low albedo of
Chiron would be due to cometary dust particles rather than irradiated
material.

In contrast, the lack of cometary activity in the Centaur Pholus is
consistent with an encompassing surface coverage by the irradiation
mantle -- witness the extreme red color and absorption features
associated with hydrocarbon materials.  Although water ice may exist
locally on the surface (Cruikshank et al. 1998), Pholus's spectral
properties are still dominated by the organic irradiated crust. If
this hypothesis is correct, cometary activity should be uncommon among
Centaurs with very red colors (irradiated material), and more common
among those with neutral (ice) colors.  As the observational sample of
Centaurs grows, this is a simple prediction that can be directly tested.
However, it remains unclear
why cometary activity was activated on Chiron and not on Pholus, even
though the two Centaurs are at similar heliocentric distances.  One
possibility is that Pholus was recently expelled from the Kuiper Belt
and has not yet been heated internally to a degree sufficient to blow
off the irradiation mantle, but this hypothesis cannot be easily
tested, given the chaotic nature of the orbits of Centaurs.

\subsubsection{Centaur and Kuiper Belt surfaces}

We summarize the spectral properties of Centaurs and KBOs in Table 3.
Thus far, water ice has been reported in 3 Centaurs (Chiron, Pholus,
1997 CU$_{26}$) and 1 KBO (1996 TO$_{66}$).  The existing data are too sparse to establish whether a correlation exists
between color and the abundance of water ice among Centaurs and KBOs. 
However, water ice is
present in all three Centaurs for which near-IR spectra are available,
and in 1 out of 3 studied KBOs.  The preponderance of water ice among
Centaurs makes us suspect that that the "low" rate of detection of
water ice in KBOs has more to do with the faintness of the targets and
the resulting low-quality spectra than with the intrinsic water
contents in KBOs.  Considering the existing data and the high cosmochemical 
abundance of water ice, we predict that water ice is ubiquitous
among {\it all} objects that originated in the Kuiper Belt, although
the amount might vary from one object to another and thus determines
the possibility for cometary activity in these bodies.  

In short, it would be a good idea to re-observe those
KBOs which show no apparent water ice feature (1993 SC and 1996
TL$_{66}$) at higher signal-to-noise ratios.  Water ice might be
present after all.

\section{Summary}

\begin{enumerate}

\item The near-infrared reflectance spectrum of Chiron is
time-variable: in 1996 and 1999 it shows an absorption feature at 2
$\micron$ due to water ice. Another absorption
feature due to water ice at 1.5 $\micron$ is also marginally detected
in 1999. The features were not present in spectra taken in 1993.

\item Chiron's time-variable spectrum is consistent with variable
dilution by the coma.  During periods of low-level outgassing, surface
features are revealed.

\item Chiron's nearly neutral spectrum suggests the surface dominance of
a dust layer created from cometary debris, consisting of
unirradiated dust particles from the interior.  Chiron's original
irradiation mantle has either been blown off or buried under this
layer.  

\item  Chiron is the third Centaur in which water ice has been
detected.  This trend suggests that water ice is common on the surface
of Centaurs. We predict that water ice is ubiquitous in {\it all}
objects originating in the Kuiper Belt.  The surface coverage of this
water ice determines its detectability.

\end{enumerate}

\noindent {\bf Note} - As we finished the preparation of this
manuscript, we received a preprint by Foster et al. (1999) in which
water ice is independently identified in spectra from 1998.  Foster et
al.  also report an unidentified absorption feature at 2.15 $\micron$
that is not confirmed in our spectra.

\noindent {\bf Acknowledgements}

\noindent The United Kingdom Infrared Telescope is operated by the
Joint Astronomy Centre on behalf of the U.K. Particle Physics and
Astronomy Research Council.

\noindent JXL thanks Ted Roush for his generosity in sharing his
software and database, Dale Cruikshank for constructive comments, and
Ronnie Hoogerwerf and Jan Kleyna for helpful discussions.  This work
was partly supported by grants to JXL and DCJ from NASA.

\newpage

\centerline{\bf FIGURE CAPTIONS}

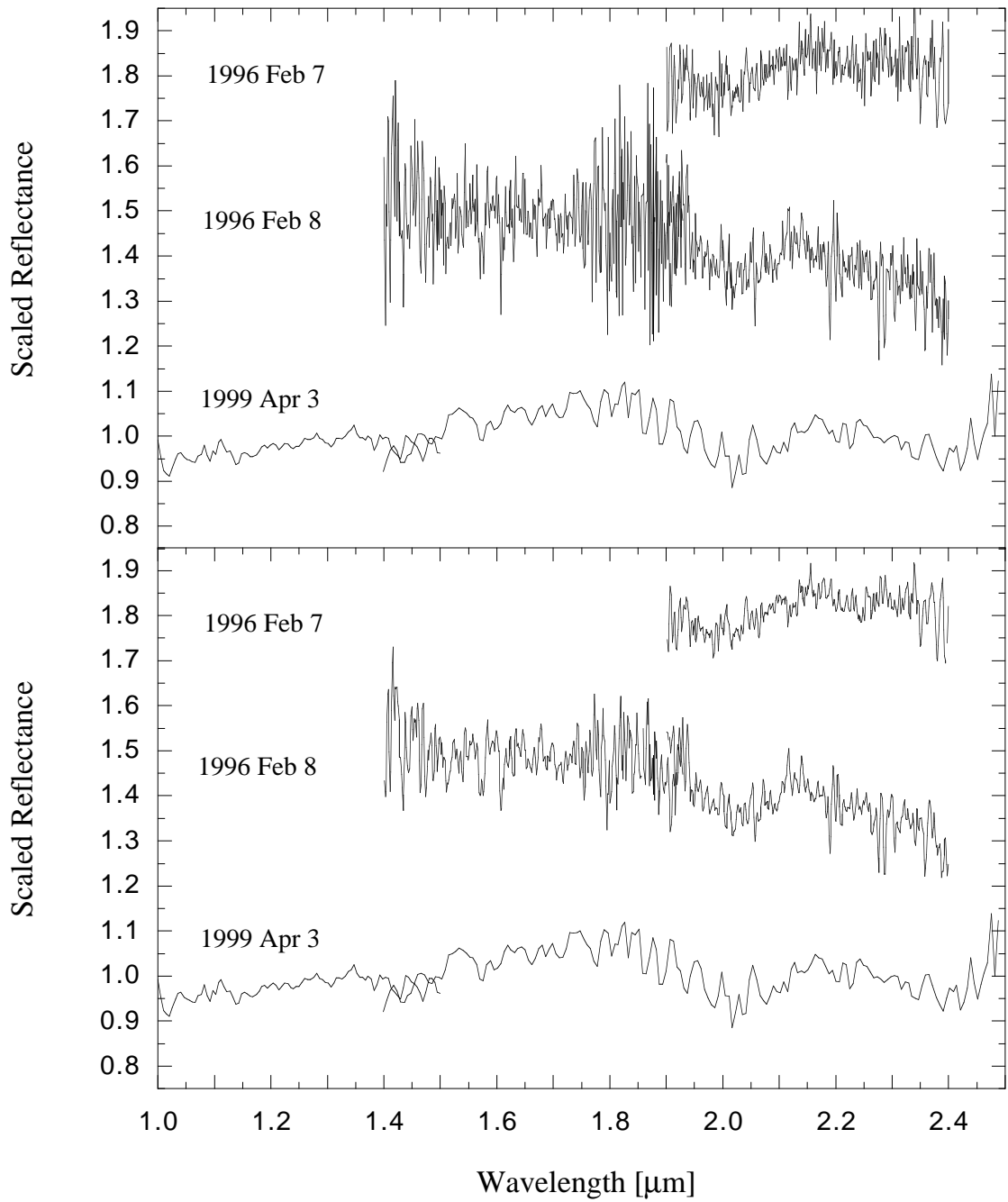
\noindent {\bf Figure 1}.  Infrared reflectance spectrum of 2060
Chiron, normalized at 2.2\ $\micron$.  The date of each spectrum
is indicated. The top panel shows the original spectra,
while in the bottom panel the 1996 spectra have been smoothed by 3
pixels (the 1999 spectrum remains unsmoothed).  There is a clear
absorption feature at 2\ $\micron$ in all three spectra, and a very
weak absorption feature at 1.5 $\micron$ in the 1999 spectra.

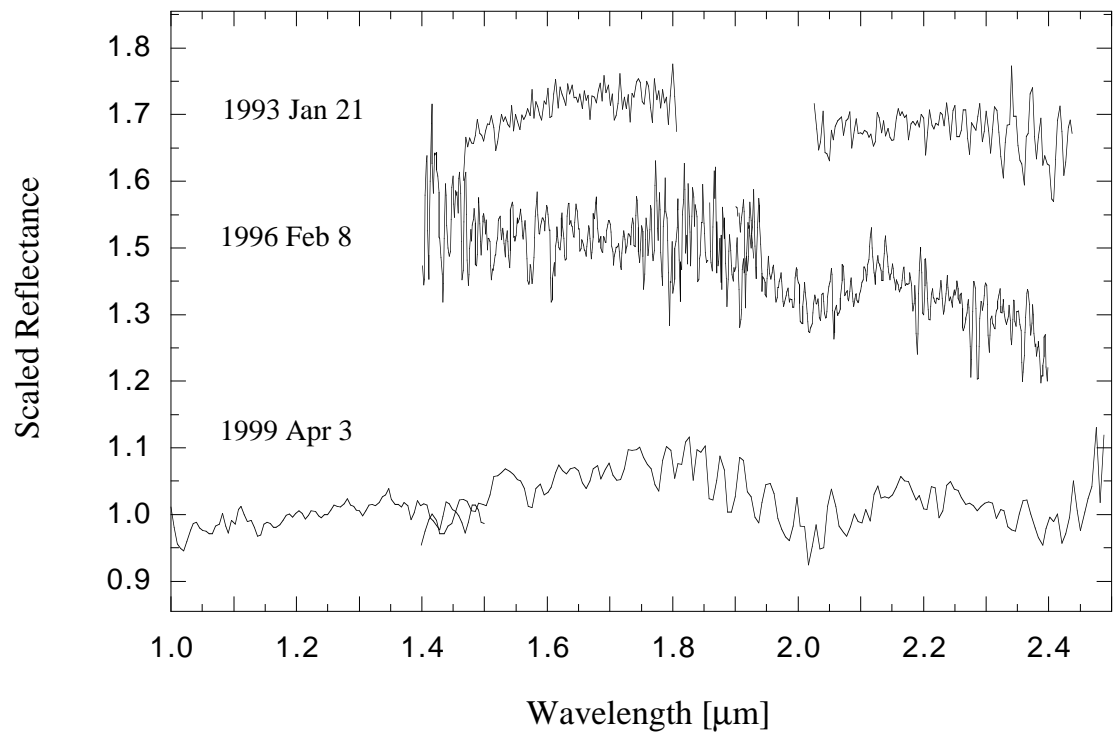
\noindent {\bf Figure 2}.  Infrared reflectance spectrum of Chiron
from 1993 (from Luu et al. 1994) compared with the 1996 and 1999
spectra.  There was no apparent spectral feature in the 1993 spectra.

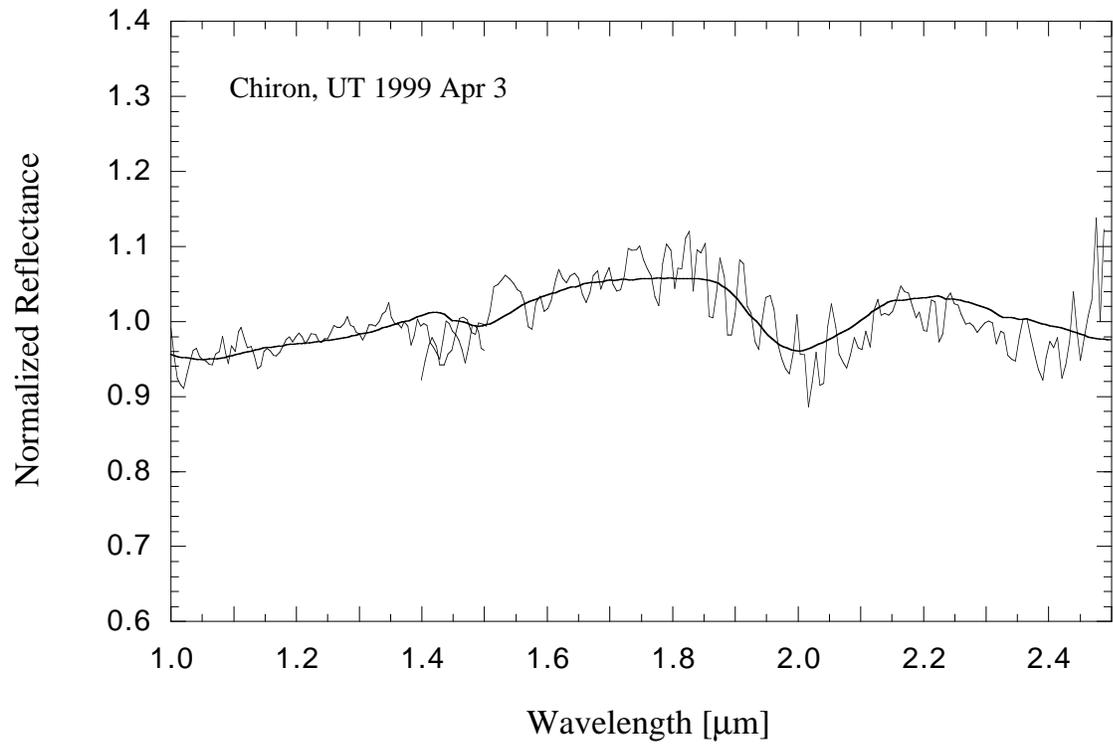
\noindent {\bf Figure 3}.  Chiron's 1996 spectra fitted with a model
consisting of a linear superposition of water ice and
olivine spectra.

\newpage

\makeatletter
\def\jnl@aj{AJ}
\ifx\revtex@jnl\jnl@aj\let\tablebreak=\nl\fi
\makeatother

\begin{deluxetable}{lllllcccc}
\footnotesize
\tablecaption{Observational Parameters of Spectra}
\tablewidth{0pc}
\tablehead{
\colhead{UT Date}&\colhead{Instrument}& 
\colhead{Wavelength}&\colhead{Slit}& 
\colhead{$\lambda / \Delta \lambda$}&
\colhead{Int\tablenotemark{a}}& R\tablenotemark{b} & 
 $\Delta$\tablenotemark{c} & $\alpha$\tablenotemark{d}\\[.2ex]
\colhead{}&\colhead{}&
\colhead{[{$\micron$}]}&\colhead{[arcsec]}&
\colhead{}&
\colhead{[sec]}&\colhead{[AU]}&\colhead{[AU]}&\colhead{[deg]}
}
\startdata

{\it UKIRT} &&&&&&&&\\
1996 Feb 7 & CGS4 K&1.3 - 2.0&1.2''$\times$ 80"&$\sim$ 850&1080&
8.45&7.86&5.6 \\
1996 Feb 8 & CGS4 H&1.3 - 2.0&1.2''$\times$ 80"&$\sim$ 850&400&
8.45&7.85&5.5 \\
1996 Feb 8 & CGS4 K&1.9 - 2.4&1.2''$\times$ 80"&$\sim$ 850&1520&
8.45&7.85&5.5 \\
{\it Keck I} &&&&&&&&\\
1999 Apr 3 & NIRC JH &1.00 - 1.55&0.68" $\times$ 38"&$\sim$ 100&600&
9.39&8.72&4.7 \\
1999 Apr 3 & NIRC HK &1.35 - 2.50&0.68" $\times$ 38"&$\sim$ 100&600&
9.39&8.72&4.7 \\
&&&&&& \\
\enddata
\tablenotetext{a}{Accumulated integration time on Chiron}
\tablenotetext{b}{Heliocentric distance}
\tablenotetext{c}{Geocentric distance}
\tablenotetext{d}{Phase angle}
\end{deluxetable}

\begin{deluxetable}{llcc}
\footnotesize
\tablecaption{Reflectivity Gradients of Chiron Spectra}
\tablewidth{0pc}
\tablehead{
\colhead{UT Date}&\colhead{Instrument}&\colhead{Wavelength Range}&
\colhead{S'}\\[.2ex]
\colhead{}&\colhead{}&\colhead{[$\micron$]}&
\colhead{[\%/1000\AA]}
}
\startdata
1996 Feb 7 & CGS4 K&2.0 -- 2.4&$0.7 \pm 0.2$\\
1996 Feb 8 & CGS4 H&1.4 -- 2.0&$-0.3 \pm 0.2$\\
1996 Feb 8 & CGS4 K&2.0 -- 2.4&$-2.1 \pm 0.3$\\

1999 Apr 3 & NIRC JH &1.0 -- 1.5&$0.9 \pm 0.2$\\
1999 Apr 3 & NIRC HK &1.5 -- 2.5&$-0.8 \pm 0.1$\\
\enddata
\end{deluxetable}

\begin{deluxetable}{llccc}
\footnotesize
\tablecaption{Spectral Properties of Centaurs and Kuiper Belt Objects}
\tablewidth{0pc}
\tablehead{
\colhead{Object}&\colhead{Type}&\colhead{$p_V$\tablenotemark{a}}&\colhead{V-J}&
\colhead{$D_{2 \micron}$\tablenotemark{b}}\\[.2ex]
\colhead{}&\colhead{}&\colhead{[\%]}&\colhead{}&
\colhead{[\%]}
}
\startdata
2060 Chiron & Centaur &$\le 0.14^{+0.06}_{-0.03}$&
$1.24 \pm 0.02$&10\\
&&(1)&(2)& This work\\
5145 Pholus & Centaur &$0.04 \pm 0.13 $&$2.59 \pm 0.02$&18\\
&&(3)&(4)&(4)\\
1997 CU$_{26}$ & Centaur &$0.04 \pm 0.01$&$1.74 \pm 0.02$&10\\
&&(5)&(6)&(7)\\
1996 TL$_{66}$& KBO &?&$1.15 \pm 0.08$&$<20$\\
&&&(8)&(9)\\
1996 TO$_{66}$ & KBO &?&$0.72 \pm 0.09$&50\\
&&&(8)&(10)\\
1993 SC & KBO &?&$1.97 \pm 0.0.08$&$<20$\\
&&&(8)&(11)\\
\enddata
\tablenotetext{a}{Geometric albedo}
\tablenotetext{b}{Depth of $2\micron$ feature}
\tablecomments{References are given in parentheses beneath each quantity:
(1) Campins et al. 1994, Altenhoff et al. 
1995, Bus et al. 1996; (2) Hartmann et al. 1990, Davies et al. 1998; 
(3) Davies et al. 1993; (4) Cruikshank et al. 1998;
(5) Jewitt and Kalas 1998; (6) McBride et al. 1999; (7) Brown and
Koresko 1998; (8) Jewitt and Luu 1998; (9) Luu and Jewitt 1998;
(10) Brown et al. 1999; (11) Brown et al. 1997.}
\end{deluxetable}

\end{document}